\documentclass[twocolumn,preprintnumbers,amsmath,amssymb,superscriptaddress]{revtex4}
\usepackage{amsmath}
\usepackage{amsfonts}
\usepackage{amssymb}
\usepackage{graphicx}
\usepackage{subfiles}
\usepackage[colorlinks,linkcolor=red,citecolor=green]{hyperref}
\usepackage{subfigure}
\usepackage{graphicx}
\usepackage{dcolumn}
\usepackage{bm}
\usepackage{amssymb,amsmath,color}
\usepackage{booktabs}
\usepackage{amsthm}
\usepackage{dsfont}
\usepackage{tcolorbox}
\usepackage{cases}
\usepackage{physics}
\usepackage{subfiles}

\newcommand{\commentold}[1]{}
\DeclareMathSymbol{:}{\mathpunct}{operators}{"3A}

\def\be{\begin{equation}}
\def\ee{\end{equation}}
\def\bea{\begin{eqnarray}}
\def\eea{\end{eqnarray}}
\def\f{\frac}
\def\n{\nonumber}
\def\l{\label}

\newcommand\numberthis{\addtocounter{equation}{1}\tag{\theequation}}

\begin{document}

\date{\today}

\def\br{\biggr}
\def\bl{\biggl}
\def\Br{\Biggr}
\def\Bl{\Biggl}
\def\be\begin{equation}
\def\ee{\end{equation}}
\def\bea{\begin{eqnarray}}
\def\eea{\end{eqnarray}}
\def\f{\frac}
\def\n{\nonumber}
\def\l{\label}

\title{Catalysis in Charging Quantum Batteries}

\author{R. R. Rodr\'iguez}
\address{International Centre for Theory of Quantum Technologies, University of Gdansk, Jana Bażyńskiego 1A, 80-309 Gdansk, Poland}
\author{B. Ahmadi}
\email{borhan.ahmadi@ug.edu.pl}
\address{International Centre for Theory of Quantum Technologies, University of Gdansk, Jana Bażyńskiego 1A, 80-309 Gdansk, Poland}
\author{P. Mazurek}
\address{International Centre for Theory of Quantum Technologies, University of Gdansk, Jana Bażyńskiego 1A, 80-309 Gdansk, Poland}
\author{S. Barzanjeh}
\address{Department of Physics and Astronomy, University of Calgary, Calgary, AB T2N 1N4 Canada}
\author{R. Alicki}
\address{International Centre for Theory of Quantum Technologies, University of Gdansk, Jana Bażyńskiego 1A, 80-309 Gdansk, Poland}
\author{P. Horodecki}
\address{International Centre for Theory of Quantum Technologies, University of Gdansk, Jana Bażyńskiego 1A, 80-309 Gdansk, Poland}

\begin{abstract}
We propose a novel approach for optimization of charging of harmonic oscillators (quantum batteries) coupled to a harmonic oscillator (charger), driven by laser field. We demonstrate that energy transfer limitations can be significantly mitigated in the presence of catalyst systems, mediating between the charger and quantum batteries. We show that these catalyst systems, either qubits or harmonic oscillators, enhance the amount of energy transferred to quantum batteries, while they themselves store almost no energy. It eliminates the need for optimizing frequency of the charging laser field, whose optimal value in the bare setting depends on coupling strengths between the charger and the batteries. 
\end{abstract}

\maketitle
\section{Introduction}
With the advent of quantum technology a growing curiosity, regarding how the current technologies can be enhanced under the principles of quantum mechanics, began. Different topics have been explored by scientists, such as quantum thermodynamics, quantum cryptography, quantum metrology and quantum computing. The quantum thermodynamics theory addresses the concepts of energy, work, heat, temperature, and entropy production in quantum systems \cite{spohn1978entropy,alicki1979quantum,parrondo2015thermodynamics,brandao2015second,ahmadi2021irreversible,ahmadi2019refined}. Recently the study of energy transfer has brought forward a diverse and rich set of phenomena, in the context of quantum batteries, to be both theoretically and experimentally investigated \cite{Alicki2013,Campaioli2017,campaioli2018quantum,barra2019dissipative,farina2019charger,julia2020bounds,crescente2022enhancing}. As in classical batteries the aim is to boost the energy transfer, from the battery to the charger, in order to be later extracted in the form of work. Many researchers have proposed and investigated possible benefits coming from quantum effects, such as quantum correlations and quantum coherence of the initial state, in charging of quantum batteries \cite{ferraro2018high,andolina2019quantum,andolina2019extractable,rossini2020quantum,kim2022operator,kim2021quantum}.

Even though this field is promising and filled with a plethora of interesting applications, there exist some issues or experimental limitations to be addressed. One of them is having the \textit{initial} quantum coherence in the state of the charger for boosting the energy transfer \cite{andolina2019extractable}. This, of course, requires specific preparation which might not be experimentally favorable. Therefore, in general, it would be preferable to start from the ground state. Quantum correlations, proposed in Ref. \cite{ferraro2018high}, may also be troublesome since uncorrelating the battery from the charger can significantly disturb the state of the battery and consumes energy. Another issue to be noticed is that it is energetically and experimentally more beneficial to use quantum harmonic oscillators (QHO), especially as a battery for storing energy, rather than two-level systems (TLS). The reason is the ability to store an \textit{arbitrary} amount of energy (up to natural technical limitations of the particular experimental setup) in one QHO while the maximum energy stored by a TLS is restricted to one quanta. The storage capacity can be increased by using many TLS though this is not experimentally desirable as it could be complicated to build or control. 

\begin{figure}[h]
\center
\includegraphics[width=8cm]{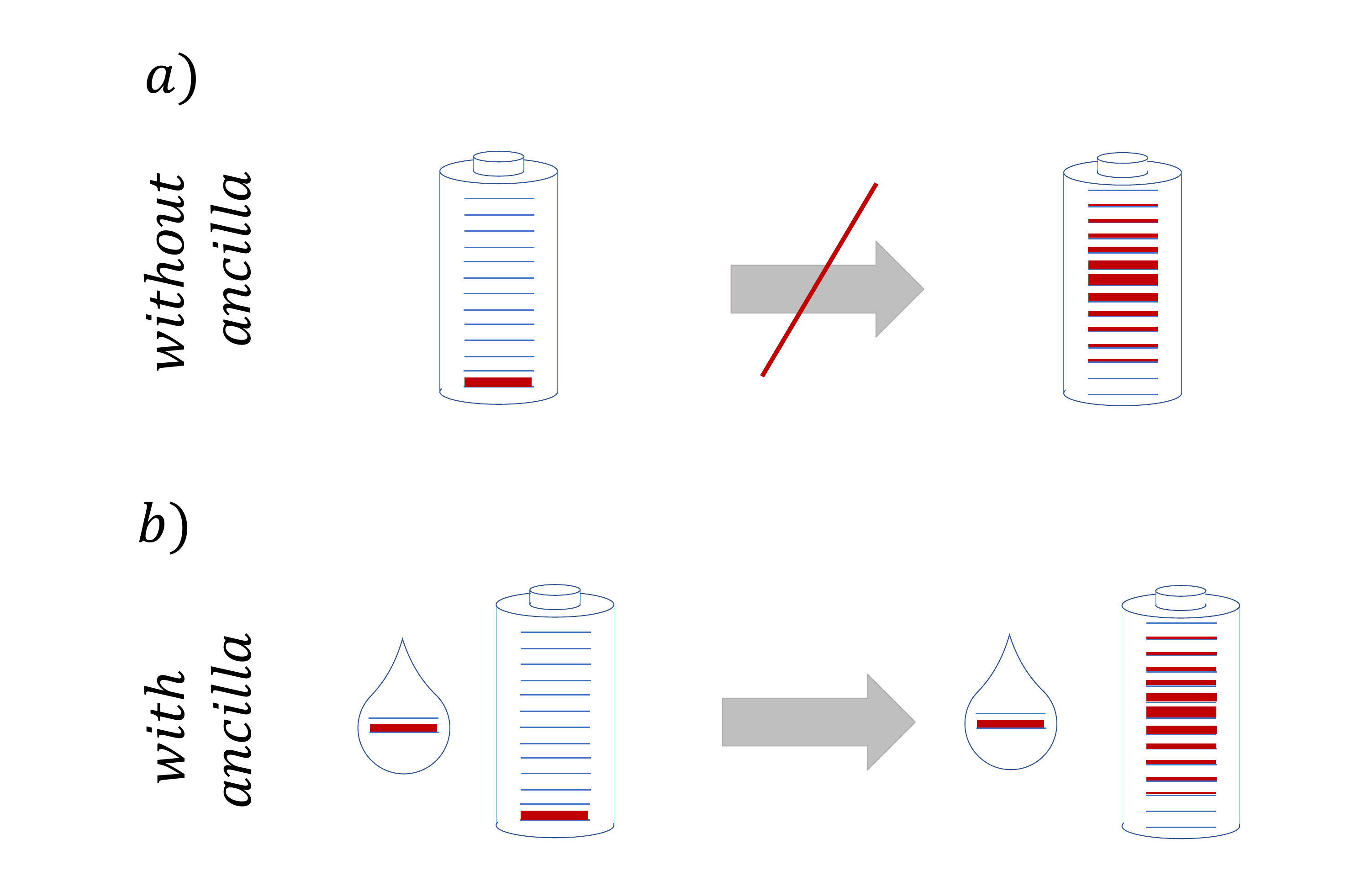}
\caption{(Color online) a) Efficient charging of a quantum system (battery) cannot be achieved by external energy driving in the presence of noise. b) Addition of a catalyst makes the energy transfer effective, while keeping the state of the catalyst almost unaffected. Note that the catalyst may be a qubit or a harmonic oscillator.}
\label{Fig1}
\end{figure}

In resource theories of entanglement and thermodynamics, tasks like entanglement distillation and state transformations can be enabled by performing them with the help of an ancillary system playing the role of a catalyst (i.e., a system whose state is not affected by the operations) \cite{jonathan1999entanglement,eisert2000catalysis}. These effects can be further enhanced if one allows for approximate catalysis, in which the state of the added system does not have to be returned completely unaltered. This may lead to trivialization of the whole theory, as arbitrary transition between states become possible. Transformations of bipartite pure entangled states without any communication, possible up to an arbitrary precision due to approximate catalysis, are an example of the so called embezzling \cite {van_Dam_2003}. \\

Motivated by these effects, here we ask about usefulness of catalylic systems outside of resource theoretical framework, namely in battery charging protocols, hoping to identify scenarios in which energy transfer is enhanced. We show that using a qubit or a harmonic oscillator ancillary system, prepared in the ground state, can boost the energy transfer in battery charging protocols, while the energy of the catalyst is left almost unaffected by the transformation. More specifically, we demonstrate that  the mere presence of a catalyst mediating between the charger and the battery can allow charger-battery systems to realize energy transfers that would otherwise require probing the coupling between the charger and the battery (see Fig. \ref{Fig1}). 

We first examine the basic model of quantum charger-battery system. As stated above, TLSs are not appropriate for storing an arbitrary amount of energy thus, to this end, we use QHOs as both quantum charger and quantum battery. An external laser field is shined on the charger and both the charger and the battery are assumed to be initially in the ground state. We then examine the intuitive way of improving energy transfer in the basic model which is based on the tuning of the external laser field with the system of the charger-battery. We will use the analysis of this case to argue for the catalytic nature of the extended setup, in which an ancillary system is added to the charger-battery platform. We then extend the method to the case of simultaneous charging of multiple batteries, in which the benefit of the catalytic approach is even more pronounced: the optimal charging laser frequency is stabilized, while in the bare setting it would depend on multiple couplings between the charger and the batteries.

While not essential for the proposed catalytic method of charging, the noise affecting it is also taken into account for illustrative purposes. Following \cite{farina2019charger}, we use the local approach in the construction of the respective master equation. We discuss the applicability of the method for any weak coupling with environment.\newline

\section{Energy transfer from a charger to a battery: The basic model}
\begin{figure}[h]
\center
\includegraphics[width=8cm]{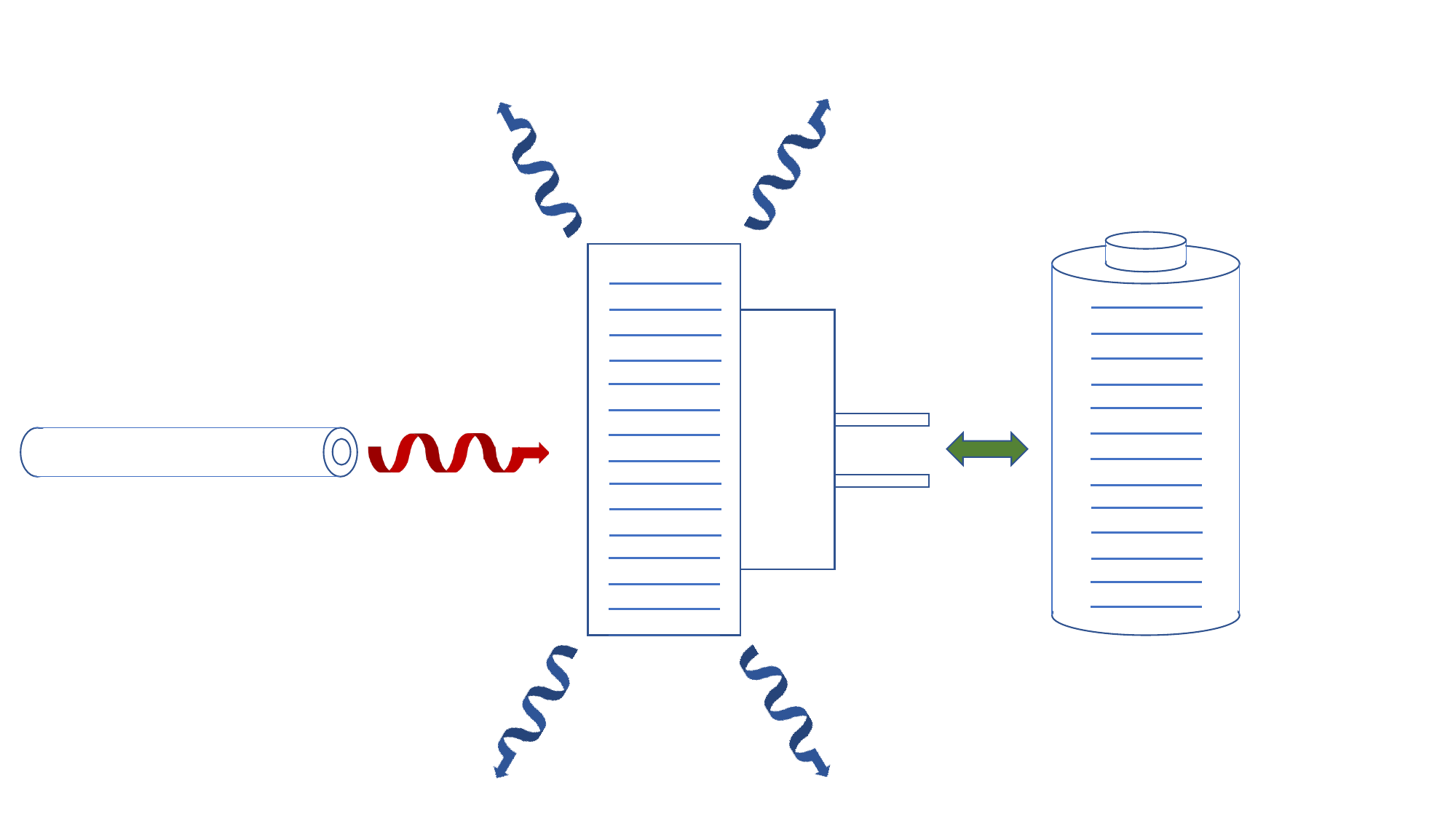}
\caption{(Color online) A QHO acting as a quantum charger (middle), powered by the classical laser field (left) and interacting with another QHO acting as a battery (right). Energy dissipation, indicated by blue curved arrows, occurs through the charger.}
\label{Fig2}
\end{figure}

In this section we present the basic model of charger-battery system and in the next sections we will present how the energy transfer can be greatly boosted in this basic model. Consider a QHO, as a charger, which interacts (charges) with another QHO (as the battery) (see Fig. \ref{Fig2})\cite{andolina2019quantum}.
A classical laser field with frequency $\omega_f$ feeds energy into the charger. Thus the Hamiltonian is written as ($\hbar=1$) 
\begin{align*}\label{Hamiltonian2}
H&=\omega_a a^\dagger a + \omega_b b^\dagger b + g(ab^\dagger + ba^\dagger)\\\nonumber
&+ F\left(e^{i\omega_ft}a + e^{-i\omega_ft}a^\dagger\right),\numberthis
\end{align*}
where $a$ ($a^\dagger$) and $b$ ($b^\dagger$) are annihilation (creation) operators of the charger $A$ and the battery $B$, respectively. Here $g$ is the coupling constant, $\omega_f$ is the laser field frequency, and $F$ the amplitude of the laser field. The interaction between the laser field and the charger has been computed in dipole approximation and rotating wave approximation (RWA) \cite{Loudon}, where the frequencies of the charger and the battery are much bigger than the decay rates, justifying the application of RWA (or secular approximation at the level of Master equation). Both the charger and the battery are prepared initially in ground states. It is assumed that the dissipation of energy into the environment $E$, occurs through the charger. Therefore the evolution of the state of the whole system $\rho_{AB}$, in interaction picture with respect to local Hamiltonians, can be described by the Lindblad form master equation \cite{Breuer}
\begin{align*}\label{me}
    \dot{\rho}_{AB}&= \mathcal{L}_{AB}[\rho_{AB}]\\\nonumber
    &= -i[g(ab^\dagger + ba^\dagger) + F\left(e^{i\Delta t}a + e^{-i\Delta t}a^\dagger\right), \rho_{AB}]\\\nonumber
    &+\gamma(N(T)+1)\mathcal{D}_a[\rho_{AB}]+\gamma N(T)\mathcal{D}_{a^\dagger}[\rho_{AB}],\numberthis
\end{align*}
where $\Delta =\omega_f-\omega_a$  is the detunning between the laser field and the charger $A$ and $N(T)=1/(e^{\omega/kT}-1)$ the average number of photon in mode $\omega$ of the environment, with Boltzman factor $k$, at temperature $T$ and
\begin{equation}\label{me}
    \mathcal{D}_c[\rho] = c\rho c^\dagger-\dfrac{1}{2}\{c^\dagger c,\rho_{AB}\},
\end{equation}
is the dissipator. The validity of the Lindblad master equation in the zero temperature requires some elaboration. In deriving the Lindblad master equation the "time-correlation" functions appear \cite{Breuer}. The width of this function represents the correlation time (memory time) $\tau_c$ and in order for the Born-Markov approximation to hold the correlation time must be short. Therefore, there might be a concern about the divergence of the correlation time as the temperature of the environment goes to zero. In Ref. \cite{gross1982superradiance} it is shown that this correlation time is equal to $\tau_c=2\pi/\omega_s$, where $\omega_s$ is the frequency of the system interacting with the bath, which clearly shows that the correlation time does not diverges. For a different argument on this issue we refer the reader to Ref. \cite{rivas2010markovian}.
\\ We assume the charger and the battery are on resonance with each other $\omega_a=\omega_b=\omega$. 
\begin{figure}[h]
\centering
\includegraphics[width=8.5cm]{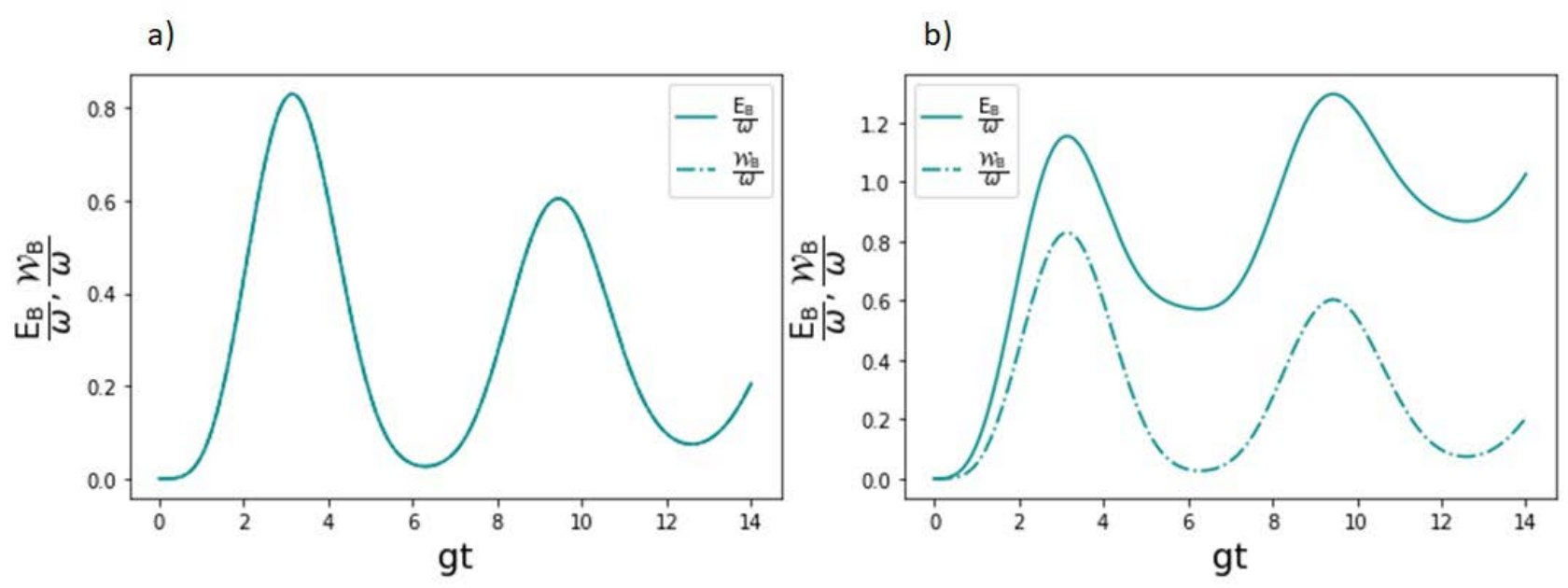}\caption{(Color online) \textbf{Off-resonant charging of a basic model.} Energy $E_B/{\omega}$ (dashed line) and ergotropy $\mathcal{W}_B/{\omega}$ (dotted line) of a quantum battery $B$, charged by a quantum charger $A$ versus $gt$ for  $g=0.2\omega$, $F=0.1\omega$, $\gamma=0.05 \omega$ and $\Delta =0$. In \textbf{a}), $N(T)=0$ and \textbf{b}), $N(T)=1$.}
\label{Fig3}
\end{figure}
The evolution can be solved exactly. For the figures of merit we choose energy stored in the battery
\begin{equation}
    E_{B}(t)=\langle b^{\dagger} b\rangle,
\end{equation}
as well as its ergotropy 
\begin{equation}
    \mathcal{W}_B=\langle b^{\dagger} b\rangle-\min_U \langle U b^{\dagger} b U^{\dagger}\rangle,
\end{equation}
quantifying the amount of work extractable from the battery by a unitary process \cite{allahverdyan2004maximal}. Energy and ergotropy of the charger, $E_{A}(t)$ and $\mathcal{W}_{A}$, are defined analogously.

For zero detuning $\Delta =0$, which is the case usually explored in the literature \cite{farina2019charger}, in Fig. \ref{Fig3} we present the time dependence of system energy and ergotropy. As is seen for zero temperature of the environment, where $N(T)=0$, the maximum energy stored in the battery is almost 0.9, i.e., less than one quanta of energy (see Fig. \ref{Fig3}\textbf{a}) and for nonzero temperature of the environment, with $N(T)=1$, the maximum energy stored in the battery is around 1.2 (see Fig. \ref{Fig3}\textbf{b}). We note that for a quantum harmonic oscillator linearly coupled to a zero-temperature bath, the Markovian dynamics maps coherent state into coherent state \cite{kossakowski1972quantum}. Since the initial state is chosen to be vacuum (also a coherent state) the state remains coherent (and hence pure) for all the times \cite{kossakowski1972quantum}. Hence the ergotropy extraction will be equal to the internal energy of the battery, i.e., all of the stored energy can be extracted in the form of work (Fig. \ref{Fig3}\textbf{a}).

In order to discuss the effect of the detuning, we look at the evolution of the energies and ergotropies of the charger and the battery \cite{farina2019charger}:
\begin{equation}\label{Wa}
    \mathcal{W}_A=\frac{16\omega F^2}{\epsilon^2}e^{-\frac{\gamma t}{2}}\sinh^2(\epsilon t/4),
\end{equation}

\begin{equation}\label{Wb}
    \mathcal{W}_B=\frac{\omega F^2}{g^2}\Big(1-e^{-\frac{\gamma t}{4}}[\cosh(\epsilon t/4)+\frac{\gamma}{\epsilon}\sinh(\epsilon t/4)]\Big)^2,
\end{equation}
where
\begin{equation}
    \epsilon=\sqrt{\gamma^2-(4g)^2}.
\end{equation}

\begin{figure}
\centering
\includegraphics[width=9cm]{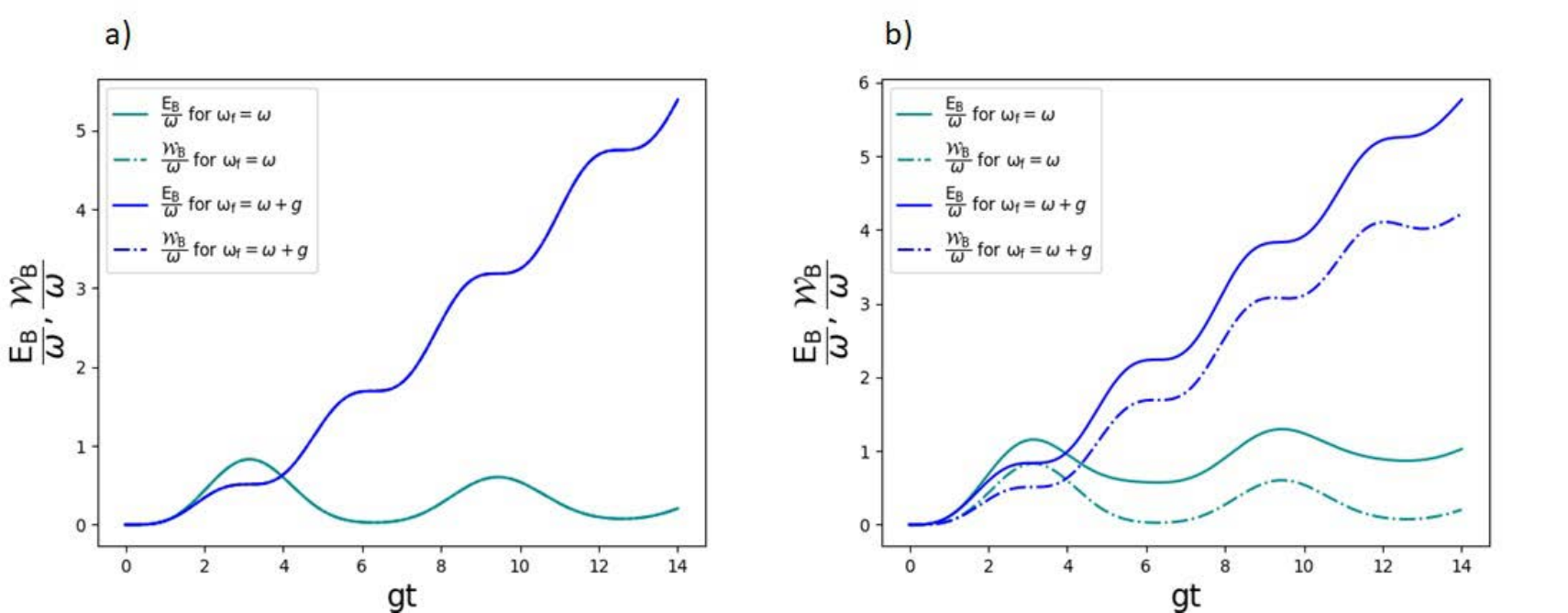}
\caption{(Color online) \textbf{On-resonant charging of a basic model. }Energy $E_B/{\omega}$ (dashed line) and ergotropy $\mathcal{W}_B/{\omega}$ (dotted line) of a quantum battery $B$, charged by a quantum charger $A$ versus $gt$ and $\omega_f=\omega_+=\omega+g$ (the same plot is obtained for $\omega_-=\omega-g$), $g=0.2\omega$, $F=0.1\omega$ and $\gamma=0.05 \omega$. In \textbf{a}) $N(T)=0$, and in \textbf{b}) $N(T)=1$. It is seen that energy transfer in the resonance case, where $\omega_f=\omega_+$, keeps increasing with time.  }
\label{Fig4}
\end{figure}

From above we see that the zero temperature solution shows limitations on the values of available ergotropy. Numerical studies (see Fig. \ref{Fig4}) reveal that this bound pertains to non-zero temperature regime. Indeed, energy stored in the charger and the battery are bounded to be of the order $(F/g)^2$, in a noiseless case oscillating with frequency of the order $g$. The effect of local noise results in suppression of these oscillations.

We point out that existence of the bound on energy stored in the charger and the battery (and present already for a noiseless case) stems entirely from the fact that the laser frequency $\omega$, while in tune with local frequencies of the charger and the battery oscillators, is \textit{out of tune} with respect to the frequency of the global charger-battery system, affected by the presence of interaction between the harmonic oscillators. Indeed, for a fixed driving amplitude $F$, non-restricted energy storing on the battery seems possible only in the limit $g\rightarrow 0$. This however requires diverging charging times. 

The off-resonance nature of the driving becomes apparent after rewriting the system Hamiltonian \eqref{Hamiltonian2}  in terms of the global super-modes operators
\begin{equation}\label{smo1}
    C_{\pm}=\frac{1}{\sqrt{2}}(a\pm b),
\end{equation}
which results in
\begin{align*}\label{HamiltonianNEW}
H&= \omega_+C_+^\dagger C_+ + \omega_-C_-^\dagger C_-+\frac{F}{\sqrt{2}}\left(e^{-i\omega_ft}C_+ + e^{i\omega_ft}C_+^\dagger\right)\\\nonumber
&+ \frac{F}{\sqrt{2}}\left(e^{-i\omega_ft}C_- + e^{i\omega_ft}C_-^\dagger\right),\numberthis
\end{align*}
where
\begin{equation}\label{supermode}
    \omega_{\pm}=\omega\pm g.
\end{equation}
Eq. \eqref{supermode} reveals that the detuning is especially pronounced in the strong-coupling regime  $g\approx \omega$, blocking the energy transfer. Energy transfer can be restored by shifting the laser frequency to $\omega_{+}$ or $\omega_{-}$, which results in unrestricted charging of the global modes $C_{+}$ or $C_{-}$, respectively. In Fig. \ref{Fig4} the energy and ergotropy of the battery are plotted for $\omega_{f}=\omega_{+}$ (the same plots for $\omega_-=\omega-g$) and it is observed that the ergotropy extraction keeps boundlessly increasing with time.\\

This intuition behind the off-resonant drive will be crucial in assessing stability of the catalyst-enhanced charging protocol with respect to fluctuations of the frequency drive (Section \ref{off-res}). Here we note that for the efficient charging to take place in time scales $t$, we need to probe $\omega_{\pm}$ (or $g$) to the precision $1/t$ in order to set the on-resonant driving frequency. Analogously, time instabilities of $\omega$ or $g$ of the order $\Delta \omega$ and $\Delta g$ restrict the effective charging times to $1/\Delta \omega$ and $1/\Delta g$, respectively. The reason is the following: the off-resonance drive 
restricts the unbounded charging of the on-resonant case (see Fig. \ref{Fig4}, blue curve) to oscillate with period $1/g$ (according to (\ref{Wa}) and (\ref{Wb})), with $g$ being the measure of the detuning with respect to the resonant drive (\ref{supermode}). This is visualised by cyan-coloured curves of Fig. \ref{Fig4}, which correspond to charging with drive of frequency $\omega$, remaining out of tune with respect to the resonant frequency $\omega\pm g$ by a term $g$.

\section{Catalysis in charging quantum batteries}\label{S3}
\subsection{The model of the catalysis-assisted charging}
\begin{figure}[h]
\center
\includegraphics[width=8cm]{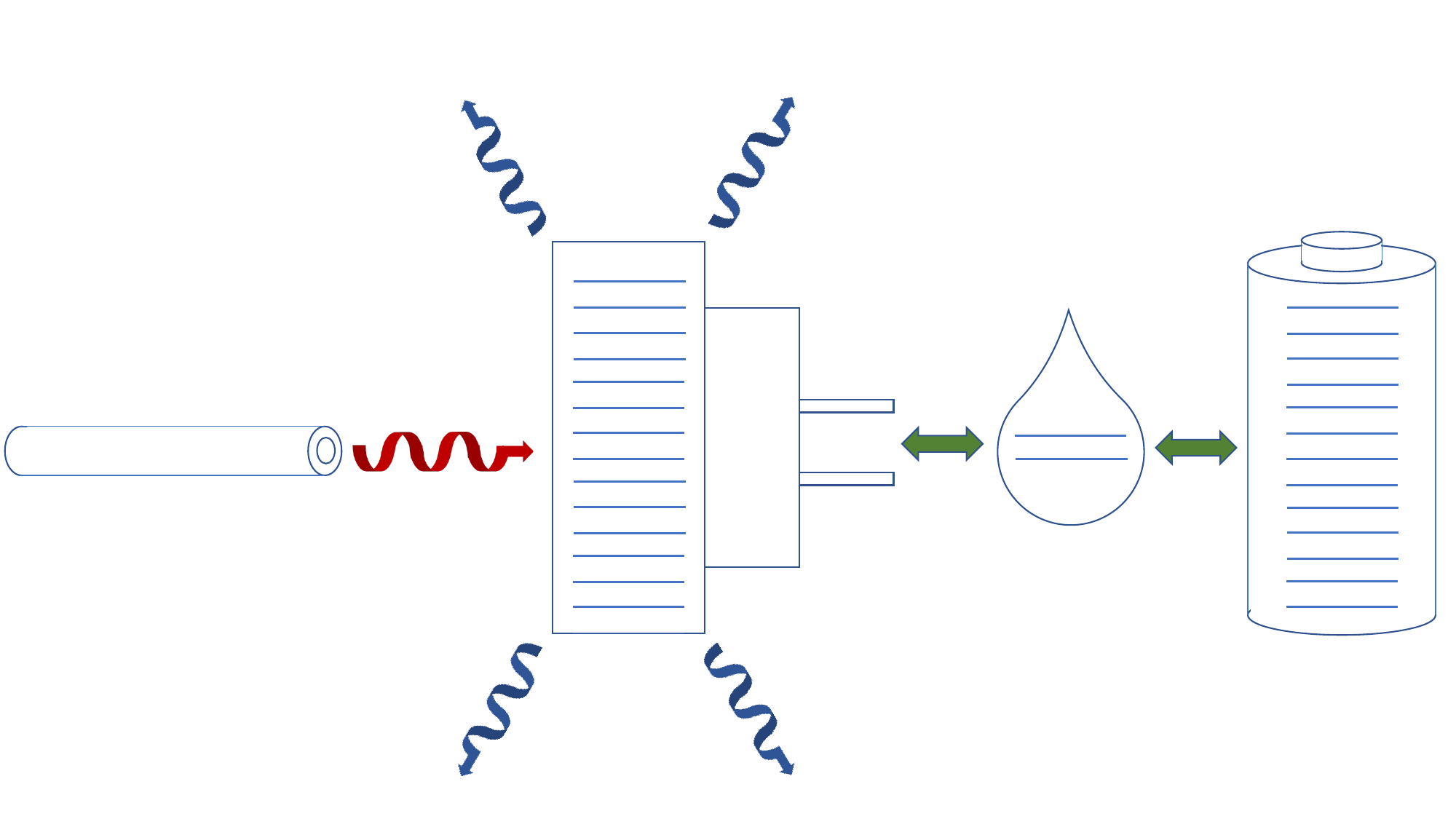}
\caption{(Color online) A two-level quantum system interacts with the charger and the battery, playing the role of a catalyst. In this paper, we investigate the usage of both qubits and harmonic oscillators as catalysts.}
\label{Fig5}
\end{figure}

In this section we propose a method for avoiding the mentioned necessity for tuning the laser frequency.
We extend the system by a catalyst mediating the interaction between the charger and the battery (see Fig. \ref{Fig5}).
The charger interacts with the catalyst and the catalyst interacts with the oscillator while there exists no direct interaction between the charger and the battery. The Hamiltonian of the whole system becomes
\begin{align*}\label{HamiltonianNEW2}
H&=\omega a^\dagger a + \omega_q q^\dagger q + \omega b^\dagger b + g_{aq}(aq^\dagger + a^\dagger q)\\\nonumber
&+ g_{bq}(bq^\dagger+b^\dagger q) + F\left(e^{i\omega_ft}a + e^{-i\omega_ft}a^\dagger\right),\numberthis
\end{align*}
where $q$ ($q^\dagger$) are either annihilation (creation) operators of the qubit ($q=\sigma_{x}-i\sigma_{y}$, with $\sigma_{x}$ and $\sigma_{y}$ denoting Pauli matrices), or of harmonic oscillator $[q,q^{\dagger}]=1$.

Introducing the super-mode operators
\begin{equation}\label{smo1}
    C_{+}=\sin{\theta}\, a+\cos{\theta}\, b,
\end{equation}
\begin{equation}\label{smo2}
    C_{-}=\cos{\theta}\, a-\sin{\theta}\, b,
\end{equation}
where
\begin{equation}\label{smo3}
    \sin{\theta}=\dfrac{g_{aq}}{\sqrt{g_{aq}^2+g_{bq}^2}},\ \quad \cos{\theta}=\dfrac{g_{bq}}{\sqrt{g_{aq}^2+g_{bq}^2}},
\end{equation}
the Hamiltonian in Eq. \eqref{Hamiltonian2} can be rewritten in the form
\begin{align*}\label{Hamiltonian3}
H&= \omega_+ C_+^\dagger C_+ + \omega_- C_-^\dagger C_- + \omega_q q^\dagger q + g(C_+q^\dagger + C_+^\dagger q)\\\nonumber
&+ F\sin{\theta}(e^{-i\omega_ft}C_+ + e^{i\omega_ft}C_+^\dagger)\\\nonumber
&+ F\cos{\theta}(e^{-i\omega_ft}C_- + e^{i\omega_ft}C_-^\dagger),\numberthis
\end{align*}
where, in this case, for the super-mode frequencies we have $\omega_{\pm}=\omega$ and $g=\sqrt{g_{aq}^2+g_{bq}^2}$. Despite zero detuning $\Delta =\omega_{f}-\omega=0$, the laser field will be on resonance with the global mode $C_{-}$ of charger-battery system (see Appendix \ref{appendix} for a generalization to arbitrary number of modes (cells) in the battery). Consequently, in the noiseless case an unlimited transfer of energy to mode $C_{-}$ takes place, independently of the value of interaction constants $g_{aq}$ and $g_{bq}$, as well as energy splitting $w_{q}$ of the catalyst, which are allowed to be unknown and fluctuate in time. In the presence of noise the energy transfer is still significantly boosted. Nevertheless, depending on the dissipation rate, a bound would appear on the energy transfer.

For the case in which the catalyst is a harmonic oscillator, we argue that it will store a negligible amount of energy. This follows from the observation that while the mode $C_{-}$ is fully decoupled from $C_{+}$ and $q$, the latter modes are charged according to the following Hamiltonian: 
\begin{align*}\label{Ham4}
H'&= \omega C_+^\dagger C_+ + \omega_q q^\dagger q+ g(C_+q^\dagger + C_+^\dagger q)\\\nonumber
&+ F\sin{\theta}(e^{-i\omega_ft}C_+ + e^{i\omega_ft}C_+^\dagger),\numberthis
\end{align*}
which, upon substitution $C_{+}\rightarrow a$,  $F\sin\theta \rightarrow F$ and $q\rightarrow b$, reproduces exactly the Hamiltonian (\ref{Hamiltonian2}) provided that we set $w_{q}=\omega$. Therefore by deciding on this value of $\omega_{q}$, we enforce a bound on the energy stored in the catalyst to be of the order of $(F\sin\theta)^2/(g_{aq}^2+g_{bq}^2)$, which in the weak driving limit is negligible. 

In the case where the catalyst is a qubit, the energy stored by it is naturally bounded by its energy splitting $\omega_{q}$. By directly solving the dynamics of the system for the qubit catalyst, below we show that its energy also remains close to 0 for the charging. Similar results can be obtained for the harmonic oscillator case.

Assuming energy splitting in the catalyst qubit be equal to $\omega_q=\omega$ and keeping $\omega = \omega_q$ in the interaction picture with respect to local Hamiltonians, the equation of motion for the system is given by:
\begin{align*}\label{me2}
    \dot{\rho}_{ABQ}&= 
     -i\big[g(aq^\dagger + qa^\dagger+ bq^\dagger + qb^\dagger)\\ &+\left(\tilde{F}a + \tilde{F}^* a^\dagger\right), \rho_{ABQ}\big]\\\nonumber
    &+\sum_{x=a,q}\gamma_{x}\Big((N(T)+1)\mathcal{D}_x[\rho_{ABQ}]\\
    &+N(T)\mathcal{D}_{x^\dagger}[\rho_{ABQ}]\Big),\numberthis
\end{align*}
with $\tilde{F}=F e^{it\Delta}$.\\

For the case of the qubit catalyst, we take $q=\sigma_x + i\sigma_y$, ($\sigma_x$ and $\sigma_y$ being Pauli matrices), and solved the master equation (\ref{me2}) numerically \cite{johansson2012qutip}. While doing so, we modelled the battery and the charger as 15-level systems, making sure that during the evolution the population on the highest levels remain negligible, assuring that finite-size effects do not appear.

On the other hand, for the harmonic-oscillator case ($[q,q^{\dagger}]=1$), the master equation (\ref{me2}) leads to equations of motion of the first and second moments of the total system. We were able to obtain an exact dynamics by solving them numerically. We present them below for the zero-temperature scenario $\gamma_{a}=\gamma$, $\gamma_{q}=0$, $N(T)=0$:

\begin{equation}\label{E1}
\langle \dot{a} \rangle = -i\tilde{F}^*-ig_{1}\langle q\rangle-\frac{\gamma}{2}\langle a\rangle
\end{equation}
\begin{equation}
\langle \dot{b} \rangle = -ig_{2}\langle q\rangle,
\end{equation}
\begin{equation}
\langle \dot{q} \rangle = -ig_{1}\langle a\rangle-ig_{2}\langle b\rangle,
\end{equation}
\begin{equation}
\langle \dot{a^{\dagger}a} \rangle = -2\Im\langle \tilde{F}a\rangle + 2g_1\text{Im}\langle a^\dagger q\rangle - \gamma\langle a^\dagger a\rangle,
\end{equation}
\begin{equation}
\langle \dot{b^{\dagger}b} \rangle = 2g_{2}\Im\langle b^{\dagger}q\rangle,
\end{equation}

\begin{equation}
\langle \dot{q^{\dagger}q} \rangle = -2g_{1}\Im \langle a^{\dagger}q\rangle-2g_{2}\Im \langle b^{\dagger}q\rangle.
\end{equation}
\begin{equation}
\langle \dot{a^{\dagger}q} \rangle = i\tilde{F}\langle q\rangle-ig_{2}\langle a^{\dagger}b\rangle-ig_{1}(\langle a^{\dagger}a\rangle-\langle q^{\dagger}q\rangle)-\frac{\gamma}{2}\langle a^{\dagger} q\rangle,
\end{equation}
\begin{equation}
\langle \dot{b^{\dagger}q} \rangle = -ig_{1}\langle a b^{\dagger}\rangle-ig_{2}\langle b^{\dagger}b\rangle+ig_{2}\langle q^{\dagger}q\rangle,
\end{equation}
\begin{equation}\label{E2}
\langle \dot{ab^{\dagger}} \rangle = -i\tilde{F}^*\langle b^{\dagger} \rangle-ig_{1}\langle b^{\dagger} q\rangle+ig_{2}\langle a q^{\dagger}\rangle-\frac{\gamma}{2}\langle ab^{\dagger}\rangle.
\end{equation}\newline

It is readily seen from the equations above that the solutions for the second-order momenta can be written as products of the solutions of the first-order momenta, i.e, for all $X, Z \in \{a, q, b, a^\dagger, q^\dagger, b^\dagger\}$ we have
\begin{equation}\label{product}
\langle XZ \rangle = \langle X\rangle\langle Z\rangle.
\end{equation}
Eq. \eqref{product} indicates that the state of the whole system is factorized throughout the entire evolution provided the initial state is a product of coherent states. Therefore, all the subsystems are in coherent states for all times, if initialized as such initially. We can write
\begin{equation}\label{coherent}
\rho_{AQB}(t) = |\alpha(t)\rangle_A\langle\alpha(t)| \otimes|\alpha'(t)\rangle_Q\langle\alpha'(t)| \otimes|\alpha''(t)\rangle_B\langle\alpha''(t)| 
\end{equation}
where $|\alpha(t)\rangle_A$, $|\alpha'(t)\rangle_Q$ and $|\alpha''(t)\rangle_B$ are coherent states of $A$, $Q$ and $B$, respectively.

\subsection{Resonant drive}
We first investigate the case of the resonant drive ($\omega_{f}=w$) of zero detuning $\Delta=0$. For catalyst qubit isolated from the environment, $\gamma_{q}=0$, numerical results presentdd in Fig. \ref{Fig6} show that the internal energy and ergotropy extraction of the battery keep increasing with time with no bound on it. This is remarkable because here the qubit acts only as a catalyst which facilitates more transfer of energy from the charger to the battery while it stores almost no energy during the whole process of charging. It also suggests that the catalyst does not get significantly entangled with neither the charger nor the battery. Consequently, resetting it for another charging process would not be connected with energy expenses. As noted above as the second moments of the equations of motion factorize, product structure of the state of the total system is preserved. As can be seen in Fig. \ref{Fig6}, the ergotropy is equal to energy also for the qubit catalyst case.  

In Fig. \ref{Fig7} the energy and ergotropy are plotted for the case when both the charger and the catalyst qubit are dissipating energy into the environment. Since the energy carried by the qubit is not significant, dissipation of energy through the qubit is not expected to affect the energy transfer. Indeed, compared to Fig. \ref{Fig6}, the results do not change considerably which confirms that the energy dissipation via the catalyst is negligible. 

\subsection{Off-resonant drive}\label{off-res}
Below we show that charging performed with the use of the catalyst shows the same type of dependence on fluctuations of the frequency of the drive, as the non-catalytic drive. Here we show the exact results obtained for the harmonic oscillator catalyst, based on eq. (\ref{E1})-(\ref{E2}). Fig. \ref{Fig8} shows the time-dependence of energies of the charger and the battery, when subjected to a drive with different offsets $\Delta/\omega$.

\begin{figure}
\centering
\includegraphics[width=9cm]{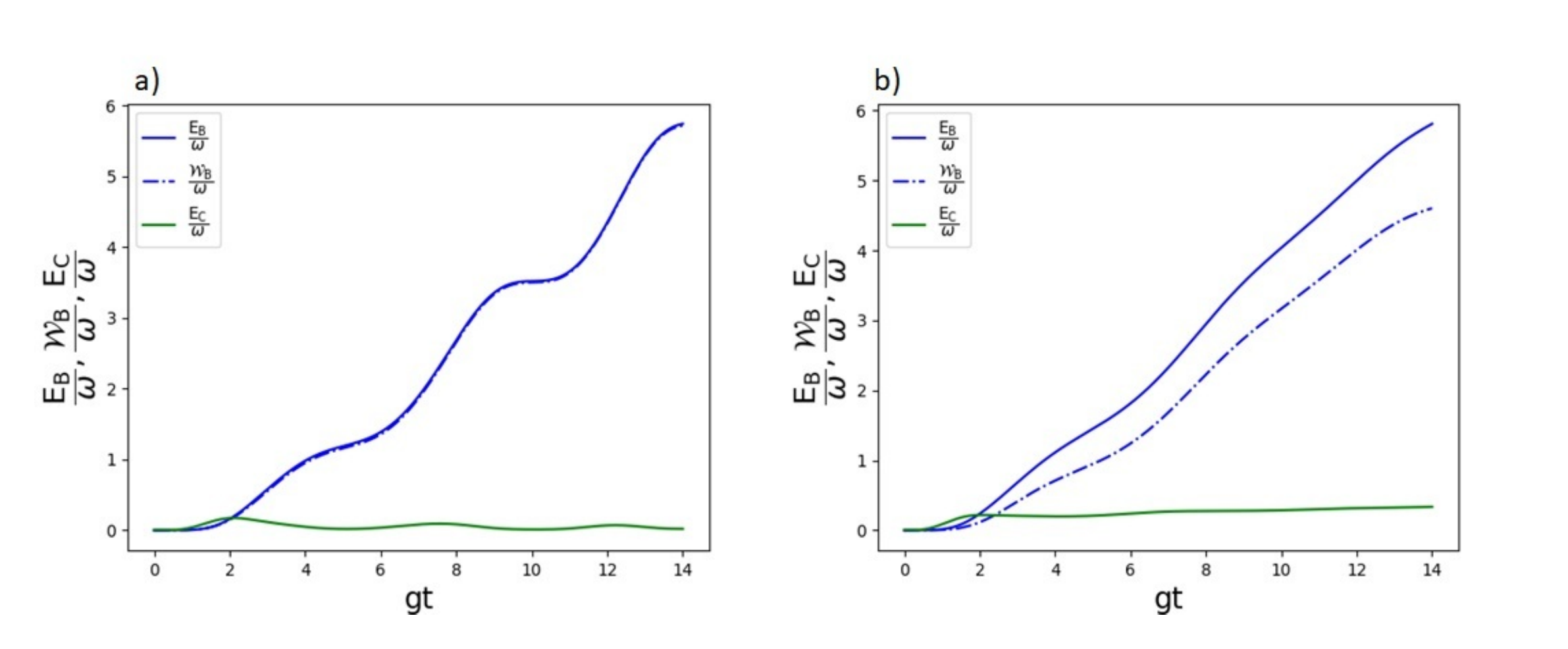}
\caption{(Color online) \textbf{Charging in a qubit catalyst model. Dissipation on the charger.} Energy $E_B/{\omega}$ (dashed line) and ergotropy $\mathcal{W}_B/{\omega}$ (dotted line) of the battery and energy $E_C(t)$ (green) of the catalyst versus $gt$ for $g_{aq}=g_{bq}=0.2\omega$, $F=0.1\omega$, $\gamma_{a}=0.05 \omega$ and $\Delta =0$. In \textbf{a)} $N(T)=0$ and in \textbf{b)} $N(T)=1$.  Adding a qubit between the charger and the battery causes the energy and the ergotropy to keep increasing with time. The catalyst qubit does not store any energy throughout the evolution.}
\label{Fig6}
\end{figure}

\begin{figure}
\centering
\includegraphics[width=6cm]{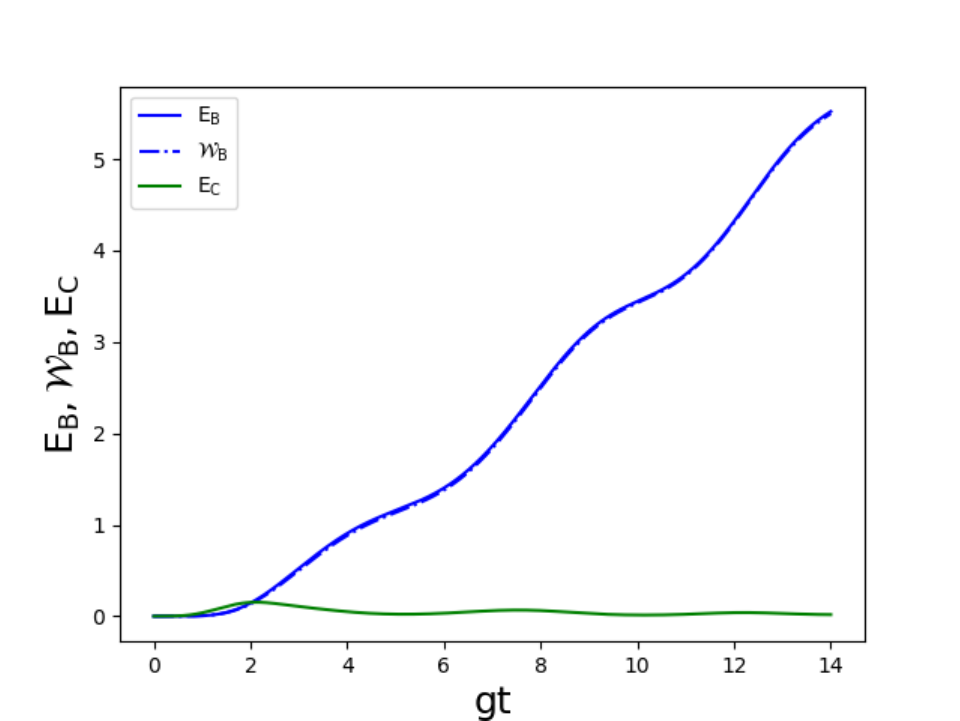}
\caption{(Color online) \textbf{Charging in a catalyst model. Dissipation on the charger and the catalyst.} Energy $E_B/{\omega}$ (dashed line) and ergotropy $\mathcal{W}_B/{\omega}$ (dotted line) of the battery and energy $E_C(t)$ (green) of the catalyst versus  $gt$ when both the charger and the catalyst qubit are dissipating energy into the vacuum for $N(T)=0$, $g_{aq}=g_{bq}=0.2\omega$, $F=0.1\omega$, $\gamma_a=\gamma_q=0.05 \omega$ and $\Delta =0$.}
\label{Fig7}
\end{figure}

\begin{figure*}
\centering
\includegraphics[width=1\textwidth]{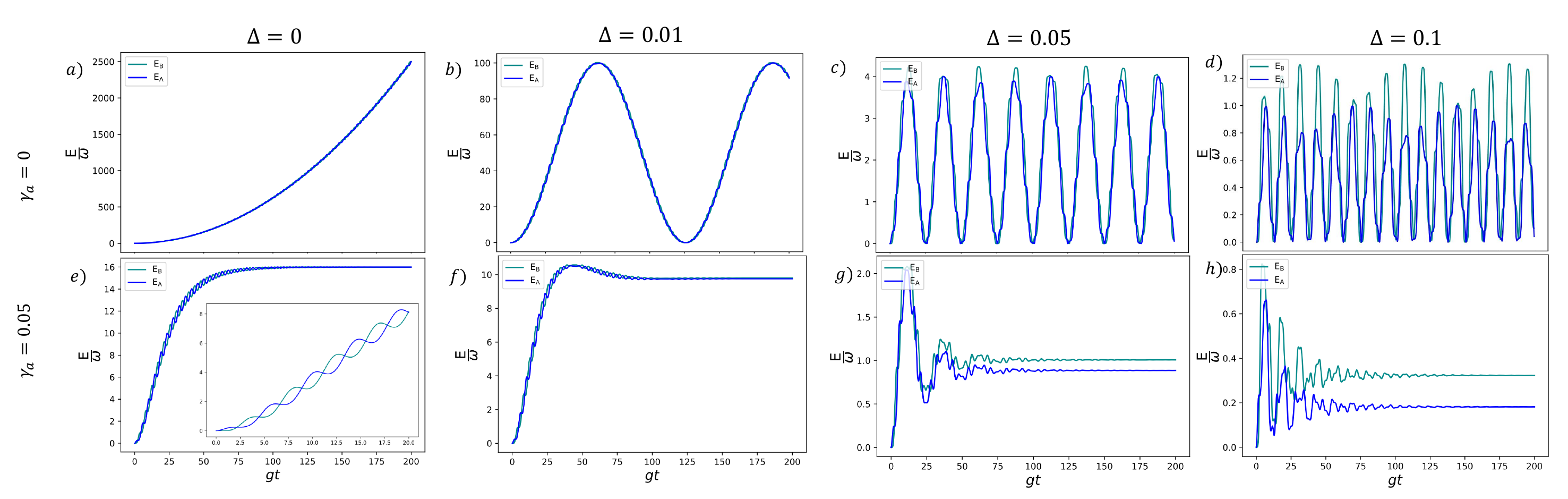}
\caption{(Color online) \textbf{Effects of detuning in battery charging assisted by harmonic oscillator catalyst.} Energies of the charger $E_{A}$ and battery $E_{B}$ for different values of local noise on the charger $\gamma_a$ and detunings $\Delta$. $N(T)=0$, $g_{aq}=g_{bq}=0.2\omega$, $F=0.1\omega$, $\gamma_b=0$. a)-d): $\gamma_a=0$; e)-h): $\gamma_a=0.05$; $\Delta=0$ for a) and e); $\Delta=0.01$ for b) and f); $\Delta=0.05$ for c) and g); $\Delta=0.1$ for d) and h). The detunings $\Delta$ are given in the units of $\omega$.}\label{Fig8}
\label{FigX}
\end{figure*}

We start the analysis by noticing that for the resonant drive in the noisy case (Fig. \ref{Fig8}e inset), the dynamics of the system with a harmonic oscillator catalyst closely resembles that containing the qubit catalyst (Fig. \ref{Fig6}) in short time scales, where it was possible to obtain the latter.  

For non-zero detunings and noisless evolution (Fig. \ref{Fig8}(a)-(d)), we see that the scaling of the period of oscillations corresponds to $1/\Delta$, in agreement with the intuition based on the analogy between the off-resonant driving of the system of two harmonic oscillators (Hamiltonian (\ref{Hamiltonian2}) and (\ref{HamiltonianNEW})) according to eq. (\ref{Wa}) and (\ref{Wb}), and that of the off-resonant driving of two virtual modes $C_{+}$ and $C_{-}$ of the system containing a catalyst ((\ref{HamiltonianNEW2}) and (\ref{Hamiltonian3})). We also observe a suppression of size of the oscillations with dependency on the detuning $(F/\Delta)^2$, as expected.

For the noisy drive (Fig. \ref{Fig8}(a)-(d)) we observe further limitations of the amplitude of the stored energy, with oscillations effectively disappearing in the overdumped regime $\gamma_{a}>\Delta$.

\begin{figure*}
\centering
\includegraphics[width=1\textwidth]{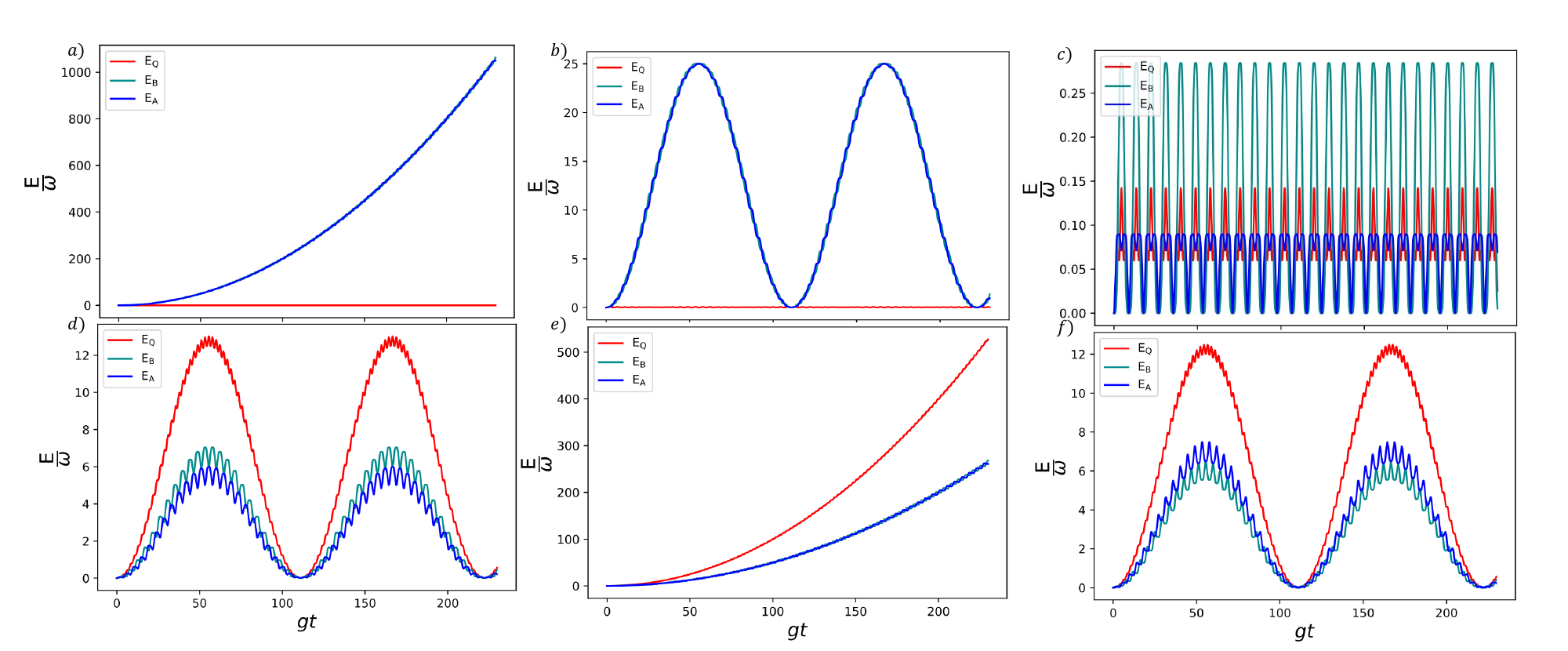}
\caption{(Color online) \textbf{Detuning with respect to different resonance frequencies.} Energies of the charger $E_{A}$, the battery $E_{B}$ and the harmonic catalyst $E_{Q}$ are plotted at and near to the resonant frequencies $\omega$ and $\omega_{+}=\omega+\sqrt{g_{aq}^2+g_{bq}^2}$. $N(T)=0$, $\gamma_a=\gamma_b=0$, $g_{aq}=g_{bq}=g=\frac{1}{2\sqrt{2}}\omega$, $F=0.1\omega$. Frequency of the drive $\omega_f$ taken to be equal to: a) $\omega$ (basic resonance, catalyst inactive); b)$1.02 \omega$;  c) $1.25 \omega$; $d) 1.48 \omega$; e) $1.5 \omega$ (global resonance, catalyst charging preferred); f) $1.52 \omega$.
}\label{Fig9}
\end{figure*}

We end the characterization of the system behaviour by focusing on the stability of the catalytic character with respect to detuning. In Fig. \ref{Fig9} we show how the detuning affects energy of the catalyst 
\begin{equation}
E_{Q}=\omega q^{\dagger}q,
\end{equation} where again we take it to be a harmonic oscillator. For better clarity of the presentation, we focus on the noisless case ($\gamma_{a}=\gamma_{b}=0$). As pointed out in the discussion of eq. (\ref{Ham4}), the system will have two more resonant frequencies, $\omega_{\pm}=\omega\pm \sqrt{g_{aq}^2+g_{bq}^2}$. In contrast to the drive with frequency $\omega$, applying a pulse with these resonant frequencies would lead to the efficient charging of the virtual modes defined on all subsystems, including the catalyst. This is what we observe from the exact numerical solution, cf. Fig. \ref{Fig9} a and Fig. \ref{Fig9} e. The period of the oscillations and the amplitude of energy of the catalyst, the charger and the battery all grow in an unbounded fashion at a resonance (notice the changing scales). 

Consequently, if the control of the driving frequency allows to set it tightly within the range $\omega\pm\sqrt{g_{aq}^2+g_{bq}^2}$ (strictly speaking, the resonant frequencies $\omega\pm\sqrt{g_{aq}^2+g_{bq}^2}$ should not be approached by a small distance shorter than $\delta \omega$ for times comparable to $1/\delta \omega$), the state of the catalyst would effectively remain unaffected by the evolution. Moreover, should the drive frequency $\omega_f$ diverge from the resonant frequency $\omega$, this would slow the charging down, but would not prevent the resumption of energy accumulation after the resonant condition is reestablished. This stems from the fact that no correlations can be built between the subsystems, as product structure (\ref{product}) is preserved also for off-resonant drive.

{\section{Experimental realisation}
 We note that the current proposal can be implemented with the current-state-of-art superconducting technology~\cite{upadhyay2021robust}. The high level control of the nano-fabrication techniques of the quantum circuits allows careful designing and fabrication of qubits and resonators with the coupling rates approaching strong-coupling regime $g_{aq}\sim\omega_q$~\cite{blais2021circuit}. The coherence time of superconducting qubits and resonators are in the order of milliseconds; allowing efficient energy exchange between the cavity and qubits. The possibility to tune the qubit-resonator coupling rate~\cite{kafri2017tunable} would allow to reach the target values. This provides additional control parameter to optimize the energy transfer between the charger and battery, and ultimately realize the current platform. 
 
 Figure \ref{Fig10}a shows the circuit of a microwave LC resonator  (such as $\lambda/2$ resonator) with inductance L and capacitance C. The Hamiltonian of the circuit can be expressed in terms of charge number $n$ and phase operator $\phi$, resulting in \cite{gross1982superradiance}
 \begin{equation}\label{QHO_ham}
     H_r=4E_C n^2-\frac{E_L}{2}\phi^2,
 \end{equation}
 where $E_C=e^2/2C$ and $E_L=(\phi_0/2\pi)^2/L$ are charging and inductive energies, respectively and $\phi_0=h/2e$. By quantizing the charge number $n= (E_L/32E_C)^{1/4} (a-a^{\dagger})$ and phase $\phi=(2E_C/E_L)^{1/4}(a+a^{\dagger})$ with $[a,a^{\dagger}]=1$, the equation (\ref{QHO_ham}) reduces to the Hamiltonian of QHO
  \begin{equation}\label{QHO_ham2}
     H_r=\hbar \omega_r (a^{\dagger}a+\frac{1}{2}),
 \end{equation}
 where $\omega_r=1/\sqrt{LC}$ is the resonance frequency of the microwave resonator. Figure \ref{Fig10}b, however, describes the circuit of a Transmon qubit with shunt capacitance $C_s$ and self-capacitance $C_J$. The Hamiltonian describing the Transmon qubit is \cite{gross1982superradiance}
 \begin{equation}
     H_q=4E_Cn^2-E_J \mathrm{cos}\phi
 \end{equation}
where the charging energy is defined by $E_C=e^2/2(C_J+C_s)$ while the Josephson energy is $E_J=I_c \phi_0/2\pi$ with $I_c$ being the critical current of the junction. By quantizing the phase and charge number and considering a sufficiently large anharmonicity in the circuit (due to strong nonlinearity in the Josephson junction), we can effectively treat the Transmon qubit as a quantum two-level system and simplifying the Hamiltonian to \cite{gross1982superradiance}
\begin{equation}
H_q=\frac{\omega_q}{2}\sigma_z
\end{equation}
where $\hbar \omega_q=\sqrt{8E_J E_C}-E_C$ is the resonance frequency of the qubit. 
\begin{figure}
\centering
\includegraphics[width=\columnwidth]{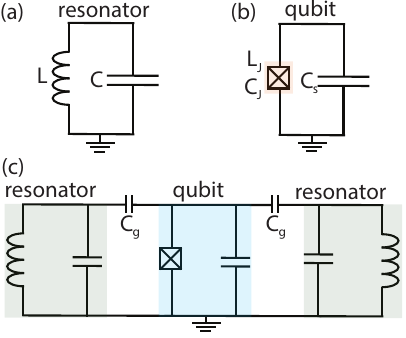}
\caption{(Color online). The electrical circuit describing: a) an LC microwave resonator with capacitance $C$ and inductance $L$, b) superconducting Transmon qubit, c) the coupling of two microwave resonators to a superconducting Transmon qubit. The resonators are coupled capacitively to the qubit using the middle capacitances $C_g$. 
}\label{Fig10}
\end{figure}
 
 The Transmon qubit can efficiently be coupled to several microwave resonators to realize battery-charger configuration described in section (\ref{S3}). Figure \ref{Fig10} (c) shows an example where two microwave resonators are capacitively coupled \cite{gross1982superradiance, blais2021circuit, barzanjeh1, barzanjeh2} to a flux tunable Transmon qubit. The coupling rates depends  on the coupling captaincies $C_g$ and can be designed to accommodate efficient energy transfer between charger and battery through the Transmon qubit. The Hamiltonian for this tripartite coupled system is \cite{gross1982superradiance, blais2021circuit}
\begin{align*}\label{Hamcircuit}
H&=\omega a^\dagger a + \omega b^\dagger b  + \omega_q \sigma_z+ g_{aq}(a + a^\dagger ) (\sigma_{-}+\sigma_{+})\\\nonumber
&+ (b+ b^\dagger ) (\sigma_{-}+\sigma_{+}) +H_{\mathrm{driv}},\numberthis
\end{align*}
where $H_{\mathrm{driv}}$ is the Hamiltonian of the drive field, $a$ and $b$ are the annihilation operators of the microwave resonators and $\sigma_i$ are the Pauli metrics describing the Transmon qubit. In the rotating wave approximation, the above Hamiltonian reduces to the battery-charger Hamiltonian (\ref{HamiltonianNEW2}).}

\section{Summary}

The use of a catalyst offers a substantial operational benefit for qubit battery charging. Here, one does not need to know the coupling strength to observe a boost in the energy transfer, which makes this setup easier to realize. On the other hand, it should be noticed that in the dispersive regime $\Delta=\omega-\omega_q\gg g$, the frequency of the catalyst is far detuned from the frequency of the charger, the energy transfer is expected to decrease and approach zero for large detunings.\\
We introduced a catalytic approach for charging quantum batteries. A catalyst initially in the ground state is added to the system and as a result the laser field, tuned to the spectrum of charger and the battery, becomes on resonance with the super-modes of the extended charger-battery-catalyst system. Consequently it is seen that the amount of energy transferred to the battery is significantly boosted, with no bound on it when there is no dissipation, while the catalyst is guaranteed to remain close to the ground state throughout the whole evolution. The method has an advantage of not relying on probing of the coupling strength, and is susceptible to its fluctuations, as well as to fluctuations of the driving frequency, as long as they remain much smaller than the coupling.

\subsection*{Acknowledgements}
We acknowledge support from the Foundation for Polish Science through IRAP project co-financed by EU within the Smart Growth Operational Program (contract no.2018/MAB/5). S.B. acknowledges funding by the Natural Sciences and Engineering Research Council of Canada (NSERC) through its Discovery Grant, funding and advisory support provided by Alberta Innovates through the Accelerating Innovations into CarE (AICE) -- Concepts Program, and support from Alberta Innovates and NSERC through Advance Grant.

\appendix
\section{Generalization to arbitrary number of modes in the battery}\label{appendix}
If we have $k$ modes (cells) in the battery and each mode interacts with the qubit separately the Hamiltonian reads
\begin{align}\label{appn1}
H&=\sum_{i=0}^{k}\omega a_i^\dagger a_i + \omega_q q^\dagger q + \sum_{i=0}^{k}g_{i}(a_iq^\dagger + a_i^\dagger q)\\\nonumber
&+ F\left(e^{-i\omega_ft}a_0 + e^{i\omega_ft}a_0^{\dagger}\right)\numberthis
\end{align}
where $a_0$ ($a^\dagger_0$) is the annihilation (creation) operator of the charger and $a_i$ ($a^\dagger_i$), $i\neq0$, are the annihilation (creation) operators of the cells of the battery. Following the Bogoljubov method \cite{Bog1958}, we define the super-mode operators $C_i$ as
\begin{equation}\label{appn2}
    C_{i}=\sum_{l=0,\dots,k}\frac{g_{il}}{n_{l}}a_{l},\ \quad i=0,\dots,k,
\end{equation}
where $n_{l}=\sqrt{\sum\limits_{j=0,\dots,k}g_{jl}^2}$, and $g_{ik}$ are constants (to be determined), such that the normalized vectors $\textbf{G}_i$ defined by $g_{ik}$ as
\begin{equation}\label{appn3}
    \textbf{G}_i=(g_{0i},\ g_{1i},\ \dots,\ g_{ki})/n_{i},\ \quad i=0,\dots,k
\end{equation}
are mutually orthogonal and
\begin{equation}\label{appn4}
    g_{0l}/n_{l}=g_{l},\  \ l=0,\dots,k.
\end{equation}
Then the Hamiltonian in Eq. \eqref{appn1} may be rewritten as
\begin{align}\label{appn5}
H&=\sum_{i=0}^{k}\omega C_i^\dagger C_i + \omega_q q^\dagger q + \left(C_0q^\dagger + C_0^\dagger q\right)\\\nonumber
&+ F\sum_{i=0}^{k}\chi_i\left(e^{-i\omega_f t}C_i + e^{i\omega_f t}C_i^\dagger\right).\numberthis
\end{align}
in which $\chi$ are functions of coefficients $g_{ik}$. In general, the transformation in Eq. (\ref{appn2}) may be described by a matrix $\hat G$ whose $i$-th column is formed by a  vector $\textbf{G}_{i}$ (note that labels of columns belong to the set $\{0,1,\dots,k\}$). We therefore have $\chi_{i}=[\hat G^{-1}]_{0i}$. 
Thus for $\Delta =0$ the laser field will be on resonance with the system of charger-battery.

\bibliography{References}

\begin{thebibliography}{35}
\expandafter\ifx\csname natexlab\endcsname\relax\def\natexlab#1{#1}\fi
\expandafter\ifx\csname bibnamefont\endcsname\relax
  \def\bibnamefont#1{#1}\fi
\expandafter\ifx\csname bibfnamefont\endcsname\relax
  \def\bibfnamefont#1{#1}\fi
\expandafter\ifx\csname citenamefont\endcsname\relax
  \def\citenamefont#1{#1}\fi
\expandafter\ifx\csname url\endcsname\relax
  \def\url#1{\texttt{#1}}\fi
\expandafter\ifx\csname urlprefix\endcsname\relax\def\urlprefix{URL }\fi
\providecommand{\bibinfo}[2]{#2}
\providecommand{\eprint}[2][]{\url{#2}}

\bibitem[{\citenamefont{Spohn}(1978)}]{spohn1978entropy}
\bibinfo{author}{\bibfnamefont{H.}~\bibnamefont{Spohn}},
  \bibinfo{journal}{Journal of Mathematical Physics}
  \textbf{\bibinfo{volume}{19}}, \bibinfo{pages}{1227} (\bibinfo{year}{1978}).

\bibitem[{\citenamefont{Alicki}(1979)}]{alicki1979quantum}
\bibinfo{author}{\bibfnamefont{R.}~\bibnamefont{Alicki}},
  \bibinfo{journal}{Journal of Physics A: Mathematical and General}
  \textbf{\bibinfo{volume}{12}}, \bibinfo{pages}{L103} (\bibinfo{year}{1979}).

\bibitem[{\citenamefont{Parrondo et~al.}(2015)\citenamefont{Parrondo, Horowitz,
  and Sagawa}}]{parrondo2015thermodynamics}
\bibinfo{author}{\bibfnamefont{J.~M.} \bibnamefont{Parrondo}},
  \bibinfo{author}{\bibfnamefont{J.~M.} \bibnamefont{Horowitz}},
  \bibnamefont{and} \bibinfo{author}{\bibfnamefont{T.}~\bibnamefont{Sagawa}},
  \bibinfo{journal}{Nature physics} \textbf{\bibinfo{volume}{11}},
  \bibinfo{pages}{131} (\bibinfo{year}{2015}).

\bibitem[{\citenamefont{Brandao et~al.}(2015)\citenamefont{Brandao, Horodecki,
  Ng, Oppenheim, and Wehner}}]{brandao2015second}
\bibinfo{author}{\bibfnamefont{F.}~\bibnamefont{Brandao}},
  \bibinfo{author}{\bibfnamefont{M.}~\bibnamefont{Horodecki}},
  \bibinfo{author}{\bibfnamefont{N.}~\bibnamefont{Ng}},
  \bibinfo{author}{\bibfnamefont{J.}~\bibnamefont{Oppenheim}},
  \bibnamefont{and} \bibinfo{author}{\bibfnamefont{S.}~\bibnamefont{Wehner}},
  \bibinfo{journal}{Proceedings of the National Academy of Sciences}
  \textbf{\bibinfo{volume}{112}}, \bibinfo{pages}{3275} (\bibinfo{year}{2015}).

\bibitem[{\citenamefont{Ahmadi et~al.}(2021)\citenamefont{Ahmadi, Salimi, and
  Khorashad}}]{ahmadi2021irreversible}
\bibinfo{author}{\bibfnamefont{B.}~\bibnamefont{Ahmadi}},
  \bibinfo{author}{\bibfnamefont{S.}~\bibnamefont{Salimi}}, \bibnamefont{and}
  \bibinfo{author}{\bibfnamefont{A.}~\bibnamefont{Khorashad}},
  \bibinfo{journal}{Scientific reports} \textbf{\bibinfo{volume}{11}},
  \bibinfo{pages}{1} (\bibinfo{year}{2021}).

\bibitem[{\citenamefont{Ahmadi et~al.}(2019)\citenamefont{Ahmadi, Salimi, and
  Khorashad}}]{ahmadi2019refined}
\bibinfo{author}{\bibfnamefont{B.}~\bibnamefont{Ahmadi}},
  \bibinfo{author}{\bibfnamefont{S.}~\bibnamefont{Salimi}}, \bibnamefont{and}
  \bibinfo{author}{\bibfnamefont{A.}~\bibnamefont{Khorashad}},
  \bibinfo{journal}{arXiv preprint arXiv:1912.01983}  (\bibinfo{year}{2019}).

\bibitem[{\citenamefont{Alicki and Fannes}(2013)}]{Alicki2013}
\bibinfo{author}{\bibfnamefont{R.}~\bibnamefont{Alicki}} \bibnamefont{and}
  \bibinfo{author}{\bibfnamefont{M.}~\bibnamefont{Fannes}},
  \bibinfo{journal}{Physical Review E} \textbf{\bibinfo{volume}{87}},
  \bibinfo{pages}{042123} (\bibinfo{year}{2013}).

\bibitem[{\citenamefont{Campaioli et~al.}(2017)\citenamefont{Campaioli,
  Pollock, Binder, Céleri, Goold, Vinjanampathy, and Modi}}]{Campaioli2017}
\bibinfo{author}{\bibfnamefont{F.}~\bibnamefont{Campaioli}},
  \bibinfo{author}{\bibfnamefont{F.~A.} \bibnamefont{Pollock}},
  \bibinfo{author}{\bibfnamefont{F.~C.} \bibnamefont{Binder}},
  \bibinfo{author}{\bibfnamefont{L.}~\bibnamefont{Céleri}},
  \bibinfo{author}{\bibfnamefont{J.}~\bibnamefont{Goold}},
  \bibinfo{author}{\bibfnamefont{S.}~\bibnamefont{Vinjanampathy}},
  \bibnamefont{and} \bibinfo{author}{\bibfnamefont{K.}~\bibnamefont{Modi}},
  \bibinfo{journal}{Physical Review Letters} \textbf{\bibinfo{volume}{118}},
  \bibinfo{pages}{150601} (\bibinfo{year}{2017}).

\bibitem[{\citenamefont{Campaioli et~al.}(2018)\citenamefont{Campaioli,
  Pollock, and Vinjanampathy}}]{campaioli2018quantum}
\bibinfo{author}{\bibfnamefont{F.}~\bibnamefont{Campaioli}},
  \bibinfo{author}{\bibfnamefont{F.~A.} \bibnamefont{Pollock}},
  \bibnamefont{and}
  \bibinfo{author}{\bibfnamefont{S.}~\bibnamefont{Vinjanampathy}},
  \emph{\bibinfo{title}{Quantum batteries - review chapter}}
  (\bibinfo{year}{2018}), \eprint{arXiv:1805.05507}.

\bibitem[{\citenamefont{Barra}(2019)}]{barra2019dissipative}
\bibinfo{author}{\bibfnamefont{F.}~\bibnamefont{Barra}},
  \bibinfo{journal}{Physical Review Letters} \textbf{\bibinfo{volume}{122}},
  \bibinfo{pages}{210601} (\bibinfo{year}{2019}).

\bibitem[{\citenamefont{Farina et~al.}(2019)\citenamefont{Farina, Andolina,
  Mari, Polini, and Giovannetti}}]{farina2019charger}
\bibinfo{author}{\bibfnamefont{D.}~\bibnamefont{Farina}},
  \bibinfo{author}{\bibfnamefont{G.~M.} \bibnamefont{Andolina}},
  \bibinfo{author}{\bibfnamefont{A.}~\bibnamefont{Mari}},
  \bibinfo{author}{\bibfnamefont{M.}~\bibnamefont{Polini}}, \bibnamefont{and}
  \bibinfo{author}{\bibfnamefont{V.}~\bibnamefont{Giovannetti}},
  \bibinfo{journal}{Physical Review B} \textbf{\bibinfo{volume}{99}},
  \bibinfo{pages}{035421} (\bibinfo{year}{2019}).

\bibitem[{\citenamefont{Juli{\`a}-Farr{\'e}
  et~al.}(2020)\citenamefont{Juli{\`a}-Farr{\'e}, Salamon, Riera, Bera, and
  Lewenstein}}]{julia2020bounds}
\bibinfo{author}{\bibfnamefont{S.}~\bibnamefont{Juli{\`a}-Farr{\'e}}},
  \bibinfo{author}{\bibfnamefont{T.}~\bibnamefont{Salamon}},
  \bibinfo{author}{\bibfnamefont{A.}~\bibnamefont{Riera}},
  \bibinfo{author}{\bibfnamefont{M.~N.} \bibnamefont{Bera}}, \bibnamefont{and}
  \bibinfo{author}{\bibfnamefont{M.}~\bibnamefont{Lewenstein}},
  \bibinfo{journal}{Physical Review Research} \textbf{\bibinfo{volume}{2}},
  \bibinfo{pages}{023113} (\bibinfo{year}{2020}).

\bibitem[{\citenamefont{Crescente et~al.}(2022)\citenamefont{Crescente,
  Ferraro, Carrega, and Sassetti}}]{crescente2022enhancing}
\bibinfo{author}{\bibfnamefont{A.}~\bibnamefont{Crescente}},
  \bibinfo{author}{\bibfnamefont{D.}~\bibnamefont{Ferraro}},
  \bibinfo{author}{\bibfnamefont{M.}~\bibnamefont{Carrega}}, \bibnamefont{and}
  \bibinfo{author}{\bibfnamefont{M.}~\bibnamefont{Sassetti}},
  \bibinfo{journal}{arXiv preprint arXiv:2202.01025}  (\bibinfo{year}{2022}).

\bibitem[{\citenamefont{Ferraro et~al.}(2018)\citenamefont{Ferraro, Campisi,
  Andolina, Pellegrini, and Polini}}]{ferraro2018high}
\bibinfo{author}{\bibfnamefont{D.}~\bibnamefont{Ferraro}},
  \bibinfo{author}{\bibfnamefont{M.}~\bibnamefont{Campisi}},
  \bibinfo{author}{\bibfnamefont{G.~M.} \bibnamefont{Andolina}},
  \bibinfo{author}{\bibfnamefont{V.}~\bibnamefont{Pellegrini}},
  \bibnamefont{and} \bibinfo{author}{\bibfnamefont{M.}~\bibnamefont{Polini}},
  \bibinfo{journal}{Physical review letters} \textbf{\bibinfo{volume}{120}},
  \bibinfo{pages}{117702} (\bibinfo{year}{2018}).

\bibitem[{\citenamefont{Andolina
  et~al.}(2019{\natexlab{a}})\citenamefont{Andolina, Keck, Mari, Giovannetti,
  and Polini}}]{andolina2019quantum}
\bibinfo{author}{\bibfnamefont{G.~M.} \bibnamefont{Andolina}},
  \bibinfo{author}{\bibfnamefont{M.}~\bibnamefont{Keck}},
  \bibinfo{author}{\bibfnamefont{A.}~\bibnamefont{Mari}},
  \bibinfo{author}{\bibfnamefont{V.}~\bibnamefont{Giovannetti}},
  \bibnamefont{and} \bibinfo{author}{\bibfnamefont{M.}~\bibnamefont{Polini}},
  \bibinfo{journal}{Physical Review B} \textbf{\bibinfo{volume}{99}},
  \bibinfo{pages}{205437} (\bibinfo{year}{2019}{\natexlab{a}}).

\bibitem[{\citenamefont{Andolina
  et~al.}(2019{\natexlab{b}})\citenamefont{Andolina, Keck, Mari, Campisi,
  Giovannetti, and Polini}}]{andolina2019extractable}
\bibinfo{author}{\bibfnamefont{G.~M.} \bibnamefont{Andolina}},
  \bibinfo{author}{\bibfnamefont{M.}~\bibnamefont{Keck}},
  \bibinfo{author}{\bibfnamefont{A.}~\bibnamefont{Mari}},
  \bibinfo{author}{\bibfnamefont{M.}~\bibnamefont{Campisi}},
  \bibinfo{author}{\bibfnamefont{V.}~\bibnamefont{Giovannetti}},
  \bibnamefont{and} \bibinfo{author}{\bibfnamefont{M.}~\bibnamefont{Polini}},
  \bibinfo{journal}{Physical review letters} \textbf{\bibinfo{volume}{122}},
  \bibinfo{pages}{047702} (\bibinfo{year}{2019}{\natexlab{b}}).

\bibitem[{\citenamefont{Rossini et~al.}(2020)\citenamefont{Rossini, Andolina,
  Rosa, Carrega, and Polini}}]{rossini2020quantum}
\bibinfo{author}{\bibfnamefont{D.}~\bibnamefont{Rossini}},
  \bibinfo{author}{\bibfnamefont{G.~M.} \bibnamefont{Andolina}},
  \bibinfo{author}{\bibfnamefont{D.}~\bibnamefont{Rosa}},
  \bibinfo{author}{\bibfnamefont{M.}~\bibnamefont{Carrega}}, \bibnamefont{and}
  \bibinfo{author}{\bibfnamefont{M.}~\bibnamefont{Polini}},
  \bibinfo{journal}{Physical Review Letters} \textbf{\bibinfo{volume}{125}},
  \bibinfo{pages}{236402} (\bibinfo{year}{2020}).

\bibitem[{\citenamefont{Kim et~al.}(2022)\citenamefont{Kim, Murugan, Olle, and
  Rosa}}]{kim2022operator}
\bibinfo{author}{\bibfnamefont{J.}~\bibnamefont{Kim}},
  \bibinfo{author}{\bibfnamefont{J.}~\bibnamefont{Murugan}},
  \bibinfo{author}{\bibfnamefont{J.}~\bibnamefont{Olle}}, \bibnamefont{and}
  \bibinfo{author}{\bibfnamefont{D.}~\bibnamefont{Rosa}},
  \bibinfo{journal}{Physical Review A} \textbf{\bibinfo{volume}{105}},
  \bibinfo{pages}{L010201} (\bibinfo{year}{2022}).

\bibitem[{\citenamefont{Kim et~al.}(2021)\citenamefont{Kim, Rosa, and
  {\v{S}}afr{\'a}nek}}]{kim2021quantum}
\bibinfo{author}{\bibfnamefont{J.}~\bibnamefont{Kim}},
  \bibinfo{author}{\bibfnamefont{D.}~\bibnamefont{Rosa}}, \bibnamefont{and}
  \bibinfo{author}{\bibfnamefont{D.}~\bibnamefont{{\v{S}}afr{\'a}nek}},
  \bibinfo{journal}{arXiv preprint arXiv:2108.02491}  (\bibinfo{year}{2021}).

\bibitem[{\citenamefont{Jonathan and Plenio}(1999)}]{jonathan1999entanglement}
\bibinfo{author}{\bibfnamefont{D.}~\bibnamefont{Jonathan}} \bibnamefont{and}
  \bibinfo{author}{\bibfnamefont{M.~B.} \bibnamefont{Plenio}},
  \bibinfo{journal}{Physical Review Letters} \textbf{\bibinfo{volume}{83}},
  \bibinfo{pages}{3566} (\bibinfo{year}{1999}).

\bibitem[{\citenamefont{Eisert and Wilkens}(2000)}]{eisert2000catalysis}
\bibinfo{author}{\bibfnamefont{J.}~\bibnamefont{Eisert}} \bibnamefont{and}
  \bibinfo{author}{\bibfnamefont{M.}~\bibnamefont{Wilkens}},
  \bibinfo{journal}{Physical Review Letters} \textbf{\bibinfo{volume}{85}},
  \bibinfo{pages}{437} (\bibinfo{year}{2000}).

\bibitem[{\citenamefont{van Dam and Hayden}(2003)}]{van_Dam_2003}
\bibinfo{author}{\bibfnamefont{W.}~\bibnamefont{van Dam}} \bibnamefont{and}
  \bibinfo{author}{\bibfnamefont{P.}~\bibnamefont{Hayden}},
  \bibinfo{journal}{Physical Review A} \textbf{\bibinfo{volume}{67}}
  (\bibinfo{year}{2003}).

\bibitem[{\citenamefont{Loudon and Petruccione}(2000)}]{Loudon}
\bibinfo{author}{\bibfnamefont{R.}~\bibnamefont{Loudon}} \bibnamefont{and}
  \bibinfo{author}{\bibfnamefont{F.}~\bibnamefont{Petruccione}},
  \emph{\bibinfo{title}{The Quantum Theory of Light}}
  (\bibinfo{publisher}{Oxford University Press}, \bibinfo{year}{2000}).

\bibitem[{\citenamefont{Breuer and Petruccione}(2002)}]{Breuer}
\bibinfo{author}{\bibfnamefont{H.~P.} \bibnamefont{Breuer}} \bibnamefont{and}
  \bibinfo{author}{\bibfnamefont{F.}~\bibnamefont{Petruccione}},
  \emph{\bibinfo{title}{The theory of open quantum systems}}
  (\bibinfo{publisher}{Oxford University Press on Demand},
  \bibinfo{year}{2002}).

\bibitem[{\citenamefont{Gross and Haroche}(1982)}]{gross1982superradiance}
\bibinfo{author}{\bibfnamefont{M.}~\bibnamefont{Gross}} \bibnamefont{and}
  \bibinfo{author}{\bibfnamefont{S.}~\bibnamefont{Haroche}},
  \bibinfo{journal}{Physics reports} \textbf{\bibinfo{volume}{93}},
  \bibinfo{pages}{301} (\bibinfo{year}{1982}).

\bibitem[{\citenamefont{Rivas et~al.}(2010)\citenamefont{Rivas, Plato, Huelga,
  and Plenio}}]{rivas2010markovian}
\bibinfo{author}{\bibfnamefont{A.}~\bibnamefont{Rivas}},
  \bibinfo{author}{\bibfnamefont{A.~D.~K.} \bibnamefont{Plato}},
  \bibinfo{author}{\bibfnamefont{S.~F.} \bibnamefont{Huelga}},
  \bibnamefont{and} \bibinfo{author}{\bibfnamefont{M.~B.}
  \bibnamefont{Plenio}}, \bibinfo{journal}{New Journal of Physics}
  \textbf{\bibinfo{volume}{12}}, \bibinfo{pages}{113032}
  (\bibinfo{year}{2010}).

\bibitem[{\citenamefont{Allahverdyan et~al.}(2004)\citenamefont{Allahverdyan,
  Balian, and Nieuwenhuizen}}]{allahverdyan2004maximal}
\bibinfo{author}{\bibfnamefont{A.~E.} \bibnamefont{Allahverdyan}},
  \bibinfo{author}{\bibfnamefont{R.}~\bibnamefont{Balian}}, \bibnamefont{and}
  \bibinfo{author}{\bibfnamefont{T.~M.} \bibnamefont{Nieuwenhuizen}},
  \bibinfo{journal}{EPL (Europhysics Letters)} \textbf{\bibinfo{volume}{67}},
  \bibinfo{pages}{565} (\bibinfo{year}{2004}).

\bibitem[{\citenamefont{Kossakowski}(1972)}]{kossakowski1972quantum}
\bibinfo{author}{\bibfnamefont{A.}~\bibnamefont{Kossakowski}},
  \bibinfo{journal}{Reports on Mathematical Physics}
  \textbf{\bibinfo{volume}{3}}, \bibinfo{pages}{247} (\bibinfo{year}{1972}).

\bibitem[{\citenamefont{Johansson et~al.}(2012)\citenamefont{Johansson, Nation,
  and Nori}}]{johansson2012qutip}
\bibinfo{author}{\bibfnamefont{J.~R.} \bibnamefont{Johansson}},
  \bibinfo{author}{\bibfnamefont{P.~D.} \bibnamefont{Nation}},
  \bibnamefont{and} \bibinfo{author}{\bibfnamefont{F.}~\bibnamefont{Nori}},
  \bibinfo{journal}{Computer Physics Communications}
  \textbf{\bibinfo{volume}{183}}, \bibinfo{pages}{1760} (\bibinfo{year}{2012}).

\bibitem[{\citenamefont{Upadhyay et~al.}(2021)\citenamefont{Upadhyay, Thomas,
  Chang, Golubev, Guthrie, Gubaydullin, Peltonen, and
  Pekola}}]{upadhyay2021robust}
\bibinfo{author}{\bibfnamefont{R.}~\bibnamefont{Upadhyay}},
  \bibinfo{author}{\bibfnamefont{G.}~\bibnamefont{Thomas}},
  \bibinfo{author}{\bibfnamefont{Y.-C.} \bibnamefont{Chang}},
  \bibinfo{author}{\bibfnamefont{D.~S.} \bibnamefont{Golubev}},
  \bibinfo{author}{\bibfnamefont{A.}~\bibnamefont{Guthrie}},
  \bibinfo{author}{\bibfnamefont{A.}~\bibnamefont{Gubaydullin}},
  \bibinfo{author}{\bibfnamefont{J.~T.} \bibnamefont{Peltonen}},
  \bibnamefont{and} \bibinfo{author}{\bibfnamefont{J.~P.}
  \bibnamefont{Pekola}}, \bibinfo{journal}{Physical Review Applied}
  \textbf{\bibinfo{volume}{16}}, \bibinfo{pages}{044045}
  (\bibinfo{year}{2021}).

\bibitem[{\citenamefont{Blais et~al.}(2021)\citenamefont{Blais, Grimsmo,
  Girvin, and Wallraff}}]{blais2021circuit}
\bibinfo{author}{\bibfnamefont{A.}~\bibnamefont{Blais}},
  \bibinfo{author}{\bibfnamefont{A.~L.} \bibnamefont{Grimsmo}},
  \bibinfo{author}{\bibfnamefont{S.}~\bibnamefont{Girvin}}, \bibnamefont{and}
  \bibinfo{author}{\bibfnamefont{A.}~\bibnamefont{Wallraff}},
  \bibinfo{journal}{Reviews of Modern Physics} \textbf{\bibinfo{volume}{93}},
  \bibinfo{pages}{025005} (\bibinfo{year}{2021}).

\bibitem[{\citenamefont{Kafri et~al.}(2017)\citenamefont{Kafri, Quintana, Chen,
  Shabani, Martinis, and Neven}}]{kafri2017tunable}
\bibinfo{author}{\bibfnamefont{D.}~\bibnamefont{Kafri}},
  \bibinfo{author}{\bibfnamefont{C.}~\bibnamefont{Quintana}},
  \bibinfo{author}{\bibfnamefont{Y.}~\bibnamefont{Chen}},
  \bibinfo{author}{\bibfnamefont{A.}~\bibnamefont{Shabani}},
  \bibinfo{author}{\bibfnamefont{J.~M.} \bibnamefont{Martinis}},
  \bibnamefont{and} \bibinfo{author}{\bibfnamefont{H.}~\bibnamefont{Neven}},
  \bibinfo{journal}{Physical Review A} \textbf{\bibinfo{volume}{95}},
  \bibinfo{pages}{052333} (\bibinfo{year}{2017}).

\bibitem[{\citenamefont{Zemlicka et~al.}(2022)\citenamefont{Zemlicka,
  Redchenko, Peruzzo, Hassani, Trioni, Barzanjeh, and Fink}}]{barzanjeh1}
\bibinfo{author}{\bibfnamefont{M.}~\bibnamefont{Zemlicka}},
  \bibinfo{author}{\bibfnamefont{E.}~\bibnamefont{Redchenko}},
  \bibinfo{author}{\bibfnamefont{M.}~\bibnamefont{Peruzzo}},
  \bibinfo{author}{\bibfnamefont{F.}~\bibnamefont{Hassani}},
  \bibinfo{author}{\bibfnamefont{A.}~\bibnamefont{Trioni}},
  \bibinfo{author}{\bibfnamefont{S.}~\bibnamefont{Barzanjeh}},
  \bibnamefont{and} \bibinfo{author}{\bibfnamefont{J.~M.} \bibnamefont{Fink}},
  \emph{\bibinfo{title}{Compact vacuum gap transmon qubits: Selective and
  sensitive probes for superconductor surface losses}} (\bibinfo{year}{2022}),
  \eprint{arXiv:2206.14104}.

\bibitem[{\citenamefont{Sett et~al.}(2022)\citenamefont{Sett, Hassani, Phan,
  Barzanjeh, Vukics, and Fink}}]{barzanjeh2}
\bibinfo{author}{\bibfnamefont{R.}~\bibnamefont{Sett}},
  \bibinfo{author}{\bibfnamefont{F.}~\bibnamefont{Hassani}},
  \bibinfo{author}{\bibfnamefont{D.}~\bibnamefont{Phan}},
  \bibinfo{author}{\bibfnamefont{S.}~\bibnamefont{Barzanjeh}},
  \bibinfo{author}{\bibfnamefont{A.}~\bibnamefont{Vukics}}, \bibnamefont{and}
  \bibinfo{author}{\bibfnamefont{J.~M.} \bibnamefont{Fink}},
  \emph{\bibinfo{title}{Emergent macroscopic bistability induced by a single
  superconducting qubit}} (\bibinfo{year}{2022}), \eprint{arXiv:2210.14182}.

\bibitem[{\citenamefont{{Bogoljubov}}(1958)}]{Bog1958}
\bibinfo{author}{\bibfnamefont{N.~N.} \bibnamefont{{Bogoljubov}}},
  \bibinfo{journal}{Il Nuovo Cimento} \textbf{\bibinfo{volume}{7}},
  \bibinfo{pages}{794} (\bibinfo{year}{1958}).

\end{thebibliography}

\end{document}